# Smartphone screens as astrometric calibrators

**Aidan Walk[a,b], Charles-Antoine Claveau[a], Michael Bottom[a], Mark Chun[a], Shane Jacobson[a], Maxwell Service[c], Jessica R. Lu[d]**

[a]Institute for Astronomy, University of Hawai'i at Mānoa, Hilo, HI
[b]Subaru Telescope, National Astronomical Observatory of Japan, Hilo, HI
[c]W. M. Keck Observatory, Waimea, HI
[d]University of California, Berkeley, CA
*Corresponding author: walk@naoj.org

## Abstract

Geometric optical distortion is a significant contributor to the astrometric error budget in large telescopes using adaptive optics. To increase astrometric precision, optical distortion calibration is necessary. We investigate using smartphone OLED screens as astrometric calibrators. Smartphones are low cost, have stable illumination, and can be quickly reconfigured to probe different spatial frequencies of an optical system's geometric distortion. In this work, we characterize the astrometric accuracy of a Samsung S20 smartphone, with a view towards providing large format, flexible astrometric calibrators for the next generation of astronomical instruments. We find the placement error of the pixels to be 189 nm ± 15 nm RMS. At this level of error, milliarcsecond astrometric accuracy can be obtained on modern astronomical instruments.

## 1. Introduction

Astrometry at the milli-arcsecond level is important for a wide range of science, including the detection of exoplanets (Wertz *et al.*, 2017), proper motion of stellar clusters (Platais *et al.,* 2018), and mass estimates of the supermassive black hole located in the center of our galaxy (Ghez *et al.,* 2008). However, this astrometric precision is fundamentally difficult to achieve because astronomical, atmospheric, and instrumental systematic errors (Schöck *et al.*, 2016, Trippe *et al.*, 2010) result in positional uncertainties well in excess of the astrometric requirements.

Adaptive optics (AO) systems can significantly reduce the primary contributor to this astrometric error, atmospheric turbulence. But the intrinsic geometric optical distortion of downstream instruments then limits the astrometric precision (Trippe *et al.*, 2010, Konopacky *et al.*, 2014, Wertz *et al.*, 2016). To mitigate this effect, a geometric optical distortion calibration is needed. There are two methods generally employed to do this. The first method, referred to as "self-calibration", is done by imaging a dense stellar field, such as a globular cluster, using a series of translational and rotational dithers of the telescope (Yelda *et al.*, 2010, Anderson *et al.* 2003). This maximizes data diversity, enabling the recovery of all modes in the optical distortion without any prior knowledge of the absolute positions of the stars in the targeted field. A limitation to this method, however, is that it must be performed on-sky, expending telescope time that could otherwise be allotted to observing.

The second technique employs a pinhole mask, where a regular grid of holes is placed at an instrument's focal plane and imaged through the system's optics. The geometric optical distortion is inferred by comparing the theoretical (mask) and distorted (image) point source patterns (see Figure 1). Provided the mask's pinhole placement error is sufficiently low, relative to the astrometric requirements, the point source grid can be assumed distortion free and errors on the mask will minimally contribute to the recovered optical distortion error. It is also possible to employ a calibration technique that is insensitive to the manufacturing errors on the mask by moving the mask small amounts with respect to the optical system (Cranney *et al.* 2022). Compared to the self-calibration method, pinhole masks require minimal observing time. However, pinhole masks are challenging and expensive to manufacture, and pinhole size and spacing must be designed to match the characteristics of the examined optical system. A further complication is achieving uniform illumination of the pinhole mask, which requires access to an independent light source dome lights or internal illumination.

State-of-the-art pinhole masks consist of a fused silica wafer with a chrome-on-nickel coating. A regular grid of holes is etched into the coating via photolithography. This process is extremely precise, with a desired manufacturing precision on the order of ~10 nm per 1 mm scale. A prototype pinhole mask was developed for the TMT NFIRAOS project, and its astrometric accuracy was characterized by Service *et al.,* (2019). The pinhole placement error of this mask was measured to be ~50 nm RMS, offering greater than ten times improvement compared to traditionally machined pinhole masks.

Previous work has demonstrated that smartphone-based scene generators have high pixel density (>40,000 independently controllable sources per $cm^2$), high dynamic range (~$10^4$), near Poisson-limited illumination stability, and no measurable background flux (Bottom *et al.,* 2018). These advantages, in addition to their rapid reconfigurability, low cost, and wide availability, have made them attractive options for demanding optical testing applications. For example, such a system was used to characterize and test the Roman coronagraph EMCCD (Morrissey *et al.,* in press), and similar designs have been adopted by other leading detector development groups (Prod'homme *et al.,* 2020, 2022).

The aforementioned qualities of smartphones, and their rapid reconfigurability through an appropriate software interface, makes them intriguing candidates for use as astrometric calibrators. For example, a smartphone-based scene projector can quickly generate "pinhole" grids of various spatial frequencies, random grids, grids of varying brightness and color (Soneira, 2020), and uniform flat fields in a matter of seconds, without any mechanical change to the optical system (Bottom *et al.,* 2018). However, their astrometric potential (eg, spatial uniformity) has never been evaluated.

The following paper compares the astrometric capability of an OLED phone screen with a next-generation photolithographic pinhole mask. §2 describes the experimental setup, phone screen illumination patterns, observations, and distortion calibration. In §3 the recovered OLED phone screen distortion, photolithographic pinhole mask distortion, and optical distortion of the experimental setup are presented. §4 discusses the final results and limitations in our method.

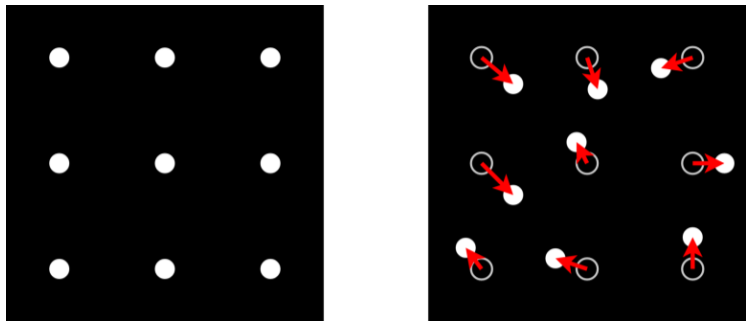

Figure 1. Left: The theoretical (mask) point source grid. Right: The point source grid after it has been imaged through an instrument's optics. The optical distortion of the instrument distorts the point source grid, as indicated by the arrows.

## 2. Methodology

### 2.1. Experimental setup

Two experimental configurations were used in this work. The first configuration was designed to verify the pinhole placement precision of a prototype pinhole mask manufactured for the TMT NFIRAOS project. This identical pinhole mask was originally calibrated by Service *et al.,* 2019, and the experimental setup is a close replication of the one detailed in that paper. The second configuration, which is a slight modification to the pinhole mask setup, is designed to measure the pixel placement error of a Samsung S20 OLED smartphone screen. Pixel placement error refers to the discrepancy between the intended positions of OLED pixels and their actual locations on the phone screen display, relative to a regular grid or other astrometric pattern. A diagram illustrating the experimental setup is shown in Figure 2.

To measure the pinhole mask, the mask is placed on a glass plate and suspended 60 mm above an optical table using a simple three-point mounting system. An OLED phone screen (serving as the illumination source) is placed beneath the pinhole mask directly laying on the optical table. A white field is displayed on the phone screen by setting every red, green, and blue diode in the pixel array (see Figure 3) to 80% peak brightness. The optical path of the system is then folded parallel to the optical table using a 45° plane mirror, and the pinhole mask is imaged through a low-distortion field lens onto a CCD camera. To achieve uniform back illumination of the pinhole mask, the phone screen is deliberately placed 60 mm away from the mask, this also ensures individual phone screen pixels are unresolved at the image plane.

The lab setup used to calibrate the OLED phone screen is nearly identical to that of the pinhole mask setup, only the pinhole mask and its mount are removed from the system. Since the pinhole mask was suspended 60 mm above the optical table, and the phone screen is placed directly on the table surface, the object plane is shifted 60 mm below where the pinhole mask was previously located. To accommodate for this shift in the object plane, the lens and camera are translated along the optical axis by the corresponding distance.

An iterative alignment procedure for the optical system has been devised and applied in order to achieve a 2f-2f configuration as accurately as possible, thus corresponding to a magnification of -1 between the object and the image projected on the CCD camera. The lens defines the optical axis, which is positioned parallel to the optical table and perpendicular to the plane of the CCD array. The normal to the mirror is set at 45 degrees from the optical axis and the table normal. The alignment quality of the optical system, as summarized at the end of this section, is estimated through the use of geometric optics.

The optical system is aligned by first locating the approximate lens and camera positions corresponding to a magnification of -1 and -2 of the phone screen[1]. The phone screen and plane mirror must not move during this step. A linear optical rail is fixed to the table and used to define the Z-axis of the optical system (see the definition of coordinate axes in Figure 2). The goal is to set the lens' optical axis parallel to the optical railing and passing through a fixed point. To do so, the center pixel of the phone screen is illuminated, which after being imaged through the 45° mirror, provides a fixed point source of light (virtually located behind the 45° mirror). This point source is imaged through the lens and forms a spot image on the camera. Then, the {lens + camera} group is translated back and forth along the rail between the magnification -1 and -2 positions. The optical axis of the lens is adjusted in translation and orientation along the (X, Y) axes. This adjustment is done to ensure that the shift of the image spot, as projected on the camera, gradually converges to zero between the two magnifications. The misalignment of the lens with respect to the optical railing can be determined by analyzing the movement of the image spot on the CCD array between the two magnification positions. This analysis applies the basic laws of conjugation for a converging lens of known focal length.

---

[1] The magnification is estimated by calculating the distance between two illuminated phone screen pixels as projected onto the CCD. This calculation assumes the average inter-pixel distance is known rather precisely (on the order of one micrometer). The adopted pixel pitch, as explained in §2.2, is 45 μm and 6 μm for the phone screen and CCD, respectively. The centroids of the image spots are determined using the method described in §2.3.

Once the lens is aligned to the optical railing by better than 1 arcminute, the CCD camera is removed from the system and replaced by a collimated laser beam. The lens is positioned one focal length away from the phone screen; therefore, the laser light comes to a focus on the phone screen. The collimated laser beam is then aligned to the approximate optical axis by making the focused laser point coincide with the illuminated pixel located at the center of the phone screen. The lens is removed from the system and a plane mirror is inserted in place of the phone. It is assumed this plane mirror lies perfectly flat on the optical table. The 45° plane mirror is aligned to the optical axis by ensuring the reflected collimated laser beam passes through an aperture at all points along the optical railing. This alignment of the 45° plane mirror is sensitive to four times the inclination of the returned collimated laser beam since the light is reflected off of the misaligned surface twice.

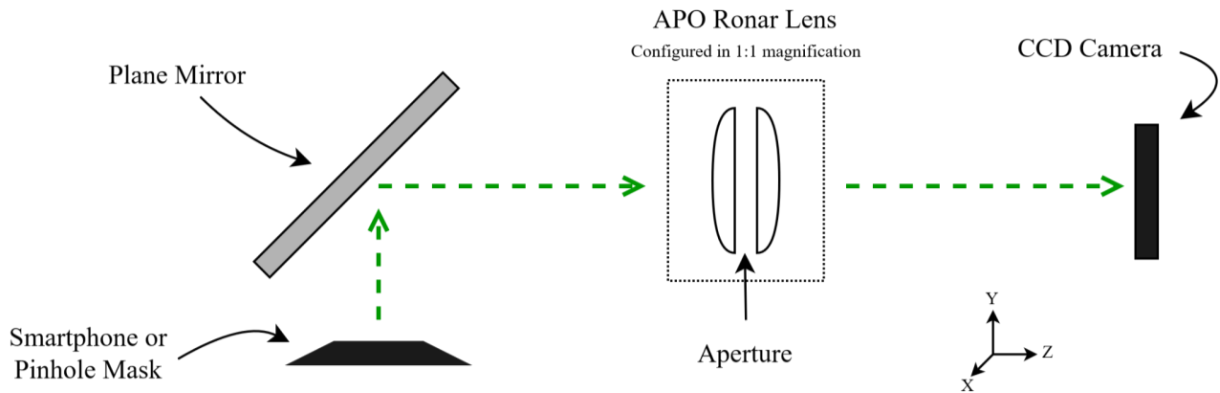

Figure 2. A diagram of the lab setup used to image the sample device (smartphone or pinhole mask). A 45° plane mirror is placed above the sample device, directing the light into a Rodenstock APO-Ronar CL 1:9 Process lens[2] with a focal length of f=480 mm and set to an f-number of f/32. The lens images the sample device onto a large-format Finger Lakes Imaging CCD Camera[3] (ML50100). The camera is 8,176 × 6,132 pixels with a pixel size of 6 μm. The dashed line indicates the optical axis of the system.

To avoid the introduction of spurious signals into the distortion measurement that might bias the result, it is important to align the surface of the astrometric mask (or phone screen) perpendicular to the optical axis. To do this, the camera is replaced by a laser point source. The lens is translated along the optical rail so that it is located approximately one focal length (f~480 mm) from the laser source. The laser beam is collimated by the lens (verified by a shearing interferometer) and aligned to the optical axis using a flat mirror lying directly on the optical table (in place of the astrometric mask or phone screen). The laser beam is aligned to the optical axis by ensuring the reflected beam coincides with the laser source. The astrometric mask is then reinstalled in the system. The incident laser beam reflects off the sample device surface and follows the reverse path through the optical system, passing once more through the 45° plane mirror and lens, and finally returns to the focus at the image plane (near where the laser point source is located). The levelness of the device can then be adjusted until the reflected laser point is once again coincident with the laser point source. In this configuration, we have sensitivity to twice the angle between the normal of the sample device surface and the collimated laser beam. Therefore, assuming a precision by eye of ~1 mm between the laser point source and the return laser spot, we can estimate an inclination of the sample device plane of ~ 5 arcmin (= $\tan^{-1}$(~ 1 [mm] / 480.) / 2.).

---

[2] https://www.rodenstock-photo.com/
[3] https://www.flicamera.com/

The camera is then reinstalled in the system and the entire alignment procedure, consisting of the 3 steps detailed above, is reiterated until no further adjustments result in significant improvements between steps. In summary, the estimated quality for the final alignment of the optical system is as follows:

Magnification = -1.0000 ± 0.0001
Lens alignment to the optical railing: X = 50 arcsec; Y = 10 arcsec
Center of the CCD array from the optical axis: X, Y = 60 μm
45° Plane Mirror: X, Y = 1 arcmin
Levelness of Astrometric Mask / Phone Screen: 5 arcmin

The final step for aligning the optical system is to fine-tune the camera focus. This is achieved by projecting a grid of pixel spots across the CCD array. The camera's focus is adjusted so as to reduce the optical field aberrations observed on the camera, playing with the lens' aperture to ensure there is no visible degradation of the point spread functions (PSFs) between different positions of the lens diaphragm, especially for the image spots located around the edge of the field of view.

## 2.2. Illumination patterns

The Samsung S20 Smartphone has a "diamond pixel" pattern ("OLED Display," n.d.; see Figure 3) with a manufacturer stated pixel density of ~220 pixels per centimeter (≃560 pixels per inch), implying a pitch of 45 μm per pixel. Because each unit cell contains four OLED pixels (one red and blue pixel, two green pixels), this pixel pitch is not representative of the actual light emitting diode size ranging from 12 μm (green pixels) to 27 μm (red pixels).

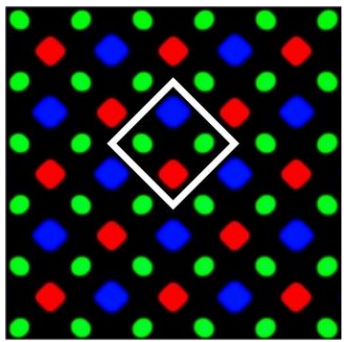

Figure 3. The diamond pentile pixel pattern of a Samsung S20 OLED phone screen. The white box denotes one unit cell and is ~45 in length and width. Each color of OLED has a different diameter: green ~12 μm, red ~27 μm, and blue ~22 μm.

The camera has a pixel size of 6 μm and the lens f-number is set to f/32. This results in a diffraction limited image of the phone screen. Therefore, the OLED pixel shapes, as shown in Figure 3, are not resolved. The phone pixels have a PSF size of ~40 μm FWHM at the image plane, or equivalently ~6.5 detector pixels (see Figure 4b). The phone screen is set to 80% peak brightness resulting in an image SNR of 500. This combination of lens f-number and SNR results in a theoretical measurement precision of 20 nm per PSF per image (Lindegren 2010). To ensure consistency between measurements, this f-number and SNR were also used while imaging the pinhole mask. Sample images of the phone screen and pinhole mask PSF's can be found in Figure 4, which shows the difference between the projected spot sizes of a green OLED pixel and a mask pinhole onto the camera detector.

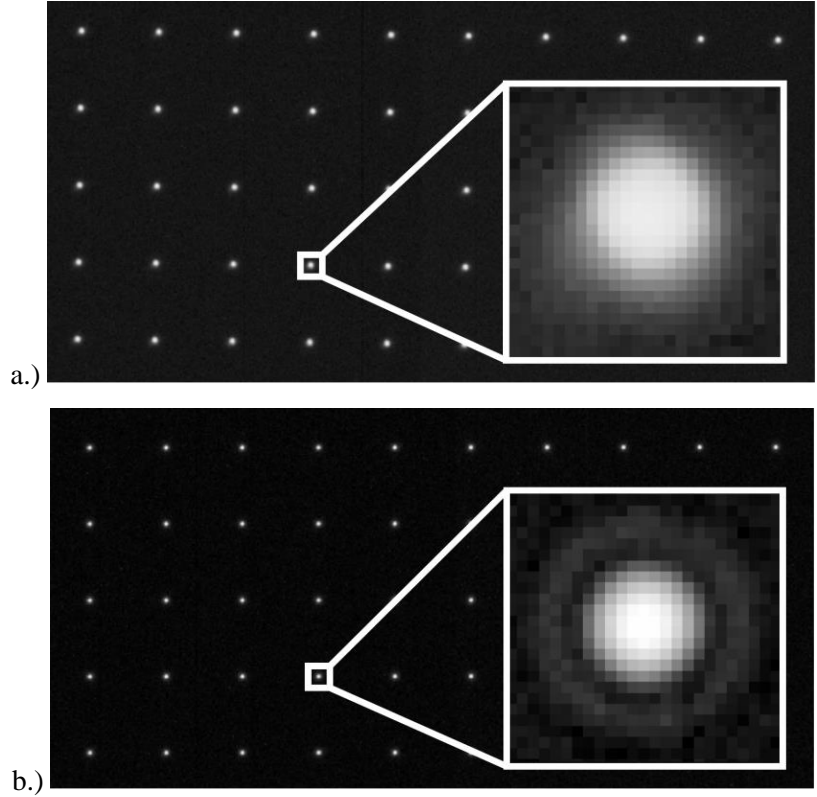

Figure 4. a.) Zoomed-in sample of the pinhole pattern on the prototype pinhole mask that was developed for the TMT NFIRAOS project. The pinhole mask is back-illuminated by an OLED phone screen displaying a white flat-field. b.) A regular grid with 1 mm spacing between illuminated pixels is displayed on the phone screen. This is a zoomed-in sample of the illumination pattern used for calibrating the pixel placement errors of the phone screen. Both images a.) and b.) are displayed in logarithmic scaling, and the zoomed-in regions (denoted by white) are taken from a 150 × 150 μm region of the camera detector.

Taking advantage of the phone screen's ability to be reprogrammed, the OLED pixel matrix can be readily reconfigured to probe different spatial frequencies in geometric optical distortion, and pixel spacing can be customized to match a wide range of optical systems. For the phone screen distortion calibration presented in this work, a 43 × 43 regular grid of pixels was illuminated on the phone screen, with a spacing of 1 mm between active pixels (as shown in Figure 4b). This illumination pattern was chosen as it replicates the TMT NFIRAOS Photolithographic Pinhole Mask pinhole pattern (Figure 4a), allowing for a more direct comparison between the two devices.

### 2.3. Distortion calibration

Due to the experimental setup containing optical distortions, both optical distortion and pixel placement error will contribute to pixel spot deviations from a regular grid at the image plane. For this reason, it would be impossible to distinguish between the optical distortion and the pixel placement error by simply imaging the phone screen in a single position with respect to the system optics. To discriminate the phone screen pixel placement errors from the system's optical distortion, we follow the methods detailed by Service *et al.*, which breaks this degeneracy by imaging the sample device in a number of different positions and angles. Any residual error in the alignment of the optical system will be preferentially captured in the optical component of the distortion solution, determined via the calibration method (i.e. without altering the measured distortion of the sample under test). This is due to the fact that the optical distortion is fixed regardless of the offset/angular position of the device.

The phone screen is imaged in a six-position dithering pattern using a combination of rotation and translation (see Figure 5 and Table 1). The phone is manually moved between each position, and 30 images are taken at each of the six phone locations. To perform source extraction, the location of pixel spots in the CCD images are first approximated using a simple threshold algorithm. Their centroids are then fit using a 2-D Gaussian function for each PSF. The resulting full dataset consists of 30 source catalogs per phone position (one source catalog for every image taken).

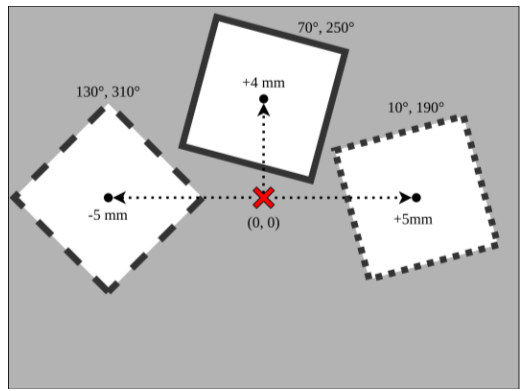

Figure 5. The Dithering pattern used to image the sample device. The grey square represents the CCD array, the center of which is denoted by an "X". The sample device, depicted as white squares, undergoes a translation across the CCD array and is imaged at two angular positions for each translation. This results in a total of six dithering positions. For example, at the white square with a dotted outline, the center of the phone screen is offset by +5 mm in the X-direction from the center of the CCD array and imaged at the 10° and 190° positions. This drawing is not to scale.

Temperature variations within the laboratory cause significant instabilities throughout the duration of an imaging period (see Figure 7), which lasts about 18 minutes per phone screen position. Therefore, before averaging the 30 images captured at each phone position, it is necessary to remove a four-parameter linear transformation from every data set. This transformation follows the same form as equation (1) but is constrained to four degrees of freedom: (X, Y) translation, rotation, and isotropic scaling. This greatly attenuates the variability in translation, rotation, and scale within the six sets of 30 images, allowing our measurement precision to improve proportionally to $\sqrt{N_{images}}$ (see Figure 6), where $N_{images}$ is the number of images averaged. Without removing this four-parameter fit from the source catalogs, measurement precision does not improve with the expected relationship. Unfortunately, this procedure precludes the recovery of the linear distortion terms in the sample device. Upon completion of this step, each stack of 30 images is averaged, resulting in six average source catalogs containing the coordinates of raw pixel spot positions, as measured on the camera detector. The final measurement precision for each average source catalog is listed in Table 1.

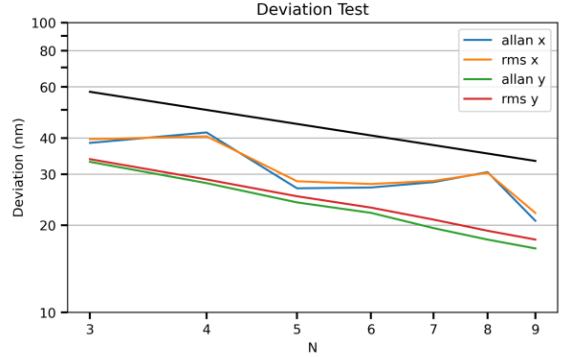

Figure 6. Allan and RMS deviations of measured spot positions for 30 sequential images taken at a single phone position. A four-parameter linear transformation has been removed from the data. The black line shows the expected $\sqrt{N_{images}}$ behavior.

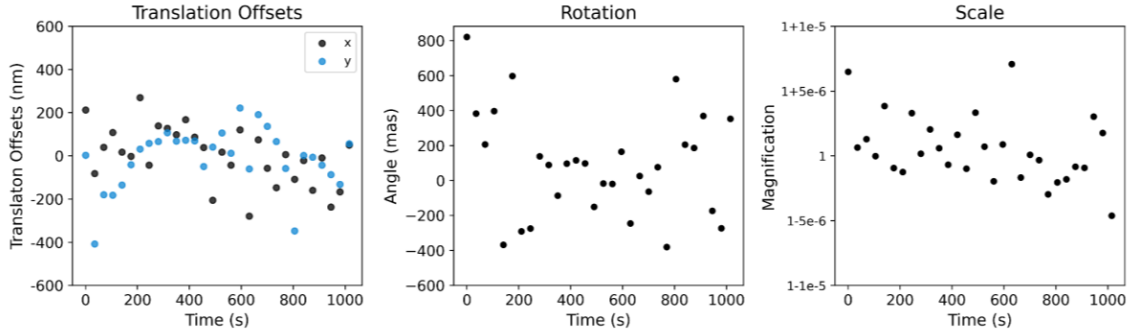

Figure 7. Bulk motion of the phone's pixel grid over an 18 minute period. Pixel movement can exceed 250 nm in translation, and a rotation of 400 milliarcseconds of the phone screen corresponds to a tangential displacement of 40 nm at the edge of the pixel grid. Scale variation is most significant, as a change in scale of $1\text{-}10^{-5}$ will contract the pixel grid by 400 nm. Equivalent motion was also observed when imaging the pinhole mask.

The six average source catalogs are then passed into a distortion recovery pipeline. This is an iterative fitting procedure, composed of three components, that solves for both the system's geometric optical distortion and phone screen's pixel placement errors. The procedure is briefly summarized below; refer to Service *et al.* paper, §4.2, for further details.

First, the detected sources are matched with a reference catalog containing idealized coordinates of the illuminated pixels, updated according to the corresponding position of the phone. These coordinates are given in an idealized distortion free camera coordinate system, which is defined with respect to the pixel array of the detector. Only the phone screen pixels present in all six catalogs are retained. The optical distortion is modeled using a six-order bivariate Legendre polynomial. The pixel spot positions are corrected based on the derived optical distortion solution, resulting in the updated distortion-free positions $(x_{cam}, y_{cam})$, still expressed in the camera coordinate frame.

Then, the linear transformation below is used to "convert" the pixel spot positions into a common reference frame (namely, the phone screen coordinate system $(x_{pix}, y_{pix})$), which corrects for the phone dither. The linear transformation can be written as follows:

$$\begin{bmatrix} x_{cam} \\ y_{cam} \end{bmatrix} = \begin{bmatrix} a_0 & a_1 & a_2 \\ b_0 & b_1 & b_2 \end{bmatrix} \begin{bmatrix} x_{pix} \\ y_{pix} \\ 1 \end{bmatrix} \quad (1)$$

As the phone is imaged in six different positions, there are six separate parameters, noted $(a_i, b_i)$, for every stacked catalog. Due to this linear transformation the Legendre polynomial (used to model the optical distortion) does not contain linear terms.

Finally, the phone screen distortion is inferred by comparing the updated pixel positions $(x_{pix}, y_{pix})$ to the reference catalog of distortion free pixel positions $(x_{ref}, y_{ref})$ following the equation below:

$$\begin{aligned} dx_{pix} &= x_{pix} - x_{ref} \\ dy_{pix} &= y_{pix} - y_{ref} \end{aligned} \quad (2)$$

Where $dx_{pix}$ and $dy_{pix}$ define the pixel placement error with respect to a regular grid given by $(x_{ref}, y_{ref})$.

In the first iteration of the fitting procedure, the deviation coordinates $(dx_{pix}, dy_{pix})$ are assumed to be zero. Therefore, the optical distortion and six linear transformations are first solved by least squares minimization, by taking $(x_{pix}, y_{pix})$ equal to $(x_{ref}, y_{ref})$. The coordinates $(x_{pix}, y_{pix})$ are then updated by applying the outcome from the previous fit to the raw pixel spot positions $(x_{cam}, y_{cam})$. These two steps are repeated four times. This procedure is designed so the positions $(x_{pix}, y_{pix})$ converge, at each iteration, to the true pixel positions of the phone screen.

Table 1. Summary of the experimental data used for recovering the phone pixel matrix distortion.

| Position | Angle [°] | $N_{exp}$ | $T_{exp}$ [s] | $N_{sources}$ | $\Delta x$ [mm] | $\Delta y$ [mm] | $\sigma_x$ [nm] | $\sigma_y$ [nm] |
|---|---|---|---|---|---|---|---|---|
| 1 | 10  | 30 | 30 | 1474 | +5 | 0  | 12.8 | 10.5 |
| 2 | 70  | 30 | 30 | 1514 | 0  | +4 | 10.2 | 10.1 |
| 3 | 130 | 30 | 30 | 1453 | -5 | 0  | 10.5 | 10.2 |
| 4 | 190 | 30 | 30 | 1492 | +5 | 0  | 11.2 | 10.1 |
| 5 | 250 | 30 | 30 | 1457 | 0  | +4 | 10.3 | 10.1 |
| 6 | 310 | 30 | 30 | 1424 | -5 | 0  | 11.0 | 10.2 |

Note. Angle is the angle of the phone screen measured in respect to the camera detector, $N_{exp}$ is the number of images taken in that phone position, $T_{exp}$ is the image exposure time in seconds, $N_{sources}$ is the number of OLED pixels visible in the image, $\Delta x$ and $\Delta y$ are the translational offset of the center of the phone screen with respect to the optical axis, and $\sigma_x$ and $\sigma_y$ are the average pixel spot measurement precisions resulting from the catalog averaging process.

The distortion recovery procedure can result in an aliasing of optical distortion into the recovered phone screen distortion, particularly when there is a need to fit large variations in scale between each measured phone position (Service *et al.* 2019.). An error in the tip/tilt of the phone screen, with respect to the system optics, will manifest as a difference in scale in the pixel matrix. As the phone screen is rotated between each reference position, this tip/tilt error becomes indecipherable from a scale variation as seen from the viewpoint of the camera.

The system is extremely sensitive to this tip/tilt error because, from the point of view of the camera, the phone screen's pixel matrix is scaled by the cosine of this error (due to the projection effect). Therefore, an error of 5 arcminutes in the levelness of the phone screen appears as a scale variation on the order of 1-

$10^{-6}$ in the pixel grid. A scale variation of this size will contract the edge of the pixel matrix by 50 nm. While this scale variation is not significant compared to the variation induced from environmental instability (Figure 7), it cannot be absorbed during the catalog averaging process. This is because the projection effect remains constant for all images taken at a single phone screen position. Consequently, this scale variation persists across all positions in the dithering pattern, resulting in an overestimation of the phone screen distortion. Future calibration systems will include a way of finely controlling the tip/tilt of the phone screen with respect to the system optics.

## 3. Results

### 3.1. Distortion solution with the phone screen

The green pixels in a Samsung S20 phone screen are illuminated with a separation of 1 mm between each pixel, spanning a total phone screen area of $43 \times 43$ mm. The phone screen is imaged in a six-position dithering pattern following Figure 5, and the data presented in Table 1 is obtained. Keeping only the pixels visible in all six phone positions, the non-linear deviations of 826 pixel positions are recovered and displayed in Figure 8. The phone screen's pixel placement error is measured to be $116 \pm 11$ nm, $150 \pm 10$ nm RMS in X and Y respectively (equal to $189 \pm 15$ nm RMS in magnitude). These deviations reflect the average pixel position error when comparing the pixel matrix to a regular reference grid.

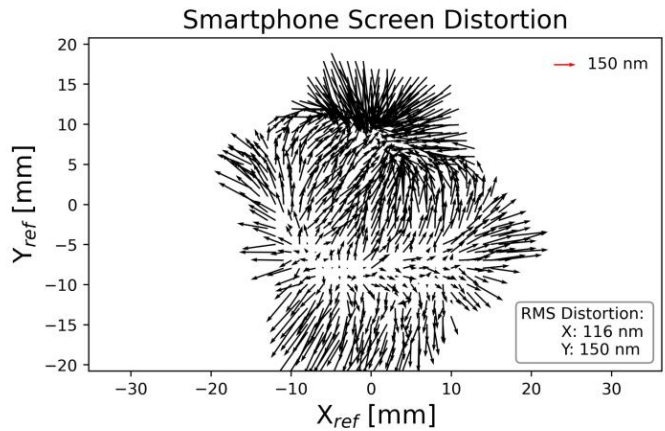

Figure 8. The positional error of 826 green OLED pixels is compared to a regular reference grid. The non-linear deviations are measured to be $189 \pm 15$ nm RMS in magnitude.

Following the procedure detailed by Service *et. al.* (2019), the system's geometric optical distortion is modeled using a six-order bivariate Legendre polynomial and measured to be $545 \pm 75$ nm RMS (see Figure 9). The geometric optical distortion solution is recovered to an accuracy of 14%, stemming from a model residual of 75 nm. This model residual reflects a combination of measurement precision and uncorrected higher order optical distortion (order > 6).

The recovered optical distortion, presented in Figure 9, shares a resemblance to the optical distortion recovered by Service *et al.* in 2019 (see Service *et al.* paper, Figure 7), albeit, with a rotation of 180°. This result is expected, as the lens used in this work (APO-Ronar CL 1:9 Process lens) is the identical lens used by Service *et al.*. Despite the structural similarity between the two recovered optical distortions, the RMS value of distortion differs by a factor of two between the two experiments. This difference likely arises from variations in the alignment of the optical system.

The linear scale factor induced by the transformation connecting the reference grid of phone pixels to the pattern of image spots projected onto the camera, after passing through the optical system, is contained in the matrix coefficients of equation 1. Its average value for the six positions of the phone screen is estimated at 1.003. This linear scale factor is a function of both the optical distortion (including any deviation from an ideal configuration of the optical system set at an exact magnification of 1) and the distortion of the tested sample device.

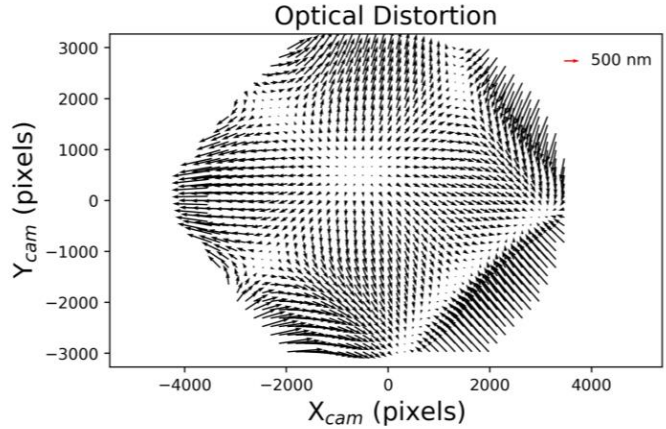

Figure 9. The non-linear geometric optical distortion of the experimental setup is modeled by a six-order bivariate Legendre polynomial and measured to be 545 ± 75 nm RMS.

To check how well we are able to recover the same RMS distortion between two realizations of the same experiment, we repeat the distortion recovery procedure five days after the results presented above. Both the recovered phone screen distortion and optical distortion are consistent within errors.

**3.2. Distortion solution with the pinhole mask**

To verify our distortion recovery of the OLED phone screen, we reproduced the results of Service *et al.*, measuring a prototype photolithographic pinhole mask manufactured for the TMT NFIRAOS project. The recovered pinhole mask distortion is displayed in Figure 10 and plotted on the same scale as the phone screen distortion displayed in Figure 8. The non-linear deviations in pinhole placement error are compared to a regular grid (i.e. pinhole mask distortion), and measured to be 53 ± 17 nm RMS.

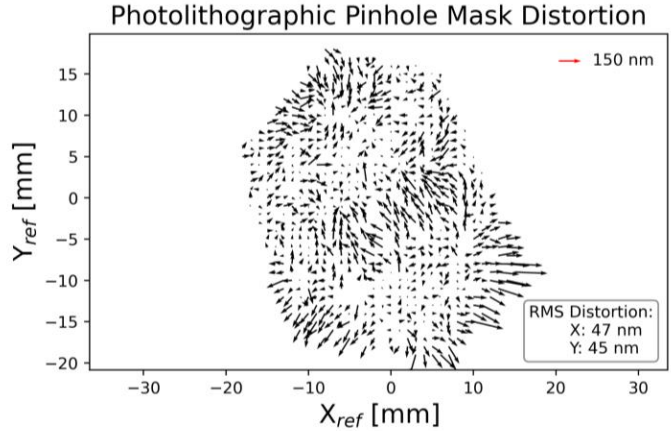

Figure 10. The pinhole placement error of the photolithographic pinhole mask developed for the TMT NFIRAOS project. The non-linear deviations of 782 pinholes are compared to a regular grid and measured to be 53 ± 17 nm RMS.

Service *et al.* paper, §4.3, reports a pinhole placement error of 36 nm, 58 nm RMS in X and Y respectively, which is a slightly different result from the calibration done in this work (shown in Figure 10). Several factors might be influencing this result, for example: environmental instabilities (such as a temperature gradient on the mask), choice of mask mounting solutions, or temporal instability in the mask distortion. However, plotting into a histogram the pinhole mask distortion measured in this work (shown in Figure 10), with the pinhole mask distortion recovered by Service *et al.*, yields two nearly indistinguishable distributions (see Figure 11). This grants us reasonable confidence in our distortion solution, particularly regarding the phone screen.

### 3.3. OLED phone screen vs. pinhole mask

The Samsung S20 OLED phone screen contains a pixel placement error ~3.5 times higher than the pinhole placement error present in the photolithographic pinhole mask. The source placement error is 189 ± 15 nm RMS for the Samsung S20 OLED phone screen, and 53 ± 17 nm RMS for the pinhole mask. These deviations reflect the non-linear distortion when comparing the sample device (pixel matrix or pinhole grid) to a perfectly regular reference grid. An overplot presented in Figure 11 shows the difference in distortion distributions between the phone screen and pinhole mask.

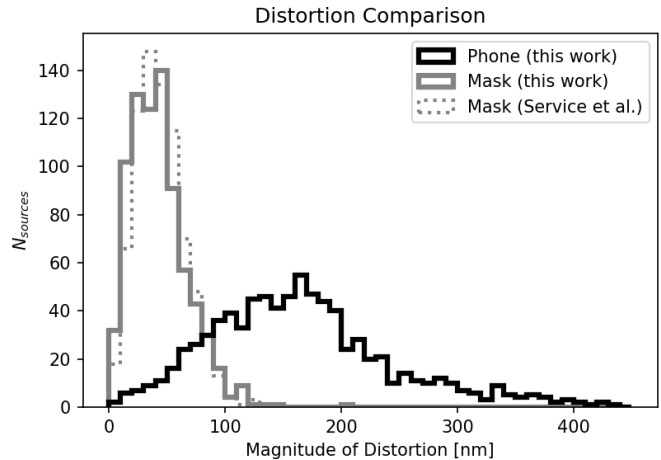

Figure 11. Phone screen distortion (black) is compared to the photolithographic pinhole mask (grey). The magnitude of distortion (i.e. $(dx^2+dy^2)^{0.5}$) for each device is plotted in a histogram. The solid and dotted grey lines represent measurements of the same pinhole mask performed three years apart. The dotted line is the result of Service *et al.* in 2019, and the solid line is our measurement of the same device performed in 2022.

## 4. Conclusion

OLED smartphones are versatile devices for applications in optical instruments. Previous research has demonstrated the utility of OLED smartphone screens for detector testing, thanks to their capacity to generate valuable calibration frames on active instruments. These frames include "pinhole" grids of various spatial frequencies, random grids, grids of varying brightness and color, and uniform flat fields (Bottom *et al.,* 2018). However, the aforementioned research solely focuses on validating the photometric capabilities of smartphone screens. This paper presents novel findings that demonstrate the effectiveness of OLED smartphone screens as astrometric calibrators.

The Samsung S20 OLED phone screen has a pixel placement error of 189 ± 15 nm RMS. These deviations reflect the non-linear pixel placement error when comparing the pixel matrix to a regular grid. This result is ~3.5 times higher than the prototype photolithographic pinhole mask developed for the TMT NFIRAOS project, and indicates OLED smartphone screens are capable astrometric calibrators for optical instruments. For instance, if a phone screen was used as an astrometric calibrator in SCExAO: VAMPIRES, employed on the Subaru Telescope (Norris *et al.,* 2015), which has an f-number of F/28.4, an astrometric precision of ~0.2 mas could be achieved. In the more general case, astronomical cameras typically range in f-number between f/2 to f/60. This correlates to an astrometric precision of 2–0.1 milliarcseconds, respectively, when using a smartphone screen as an astrometric calibrator in an 8–10 meter class telescope.

Display technologies are constantly improving, with innovations mainly stimulated by the consumer electronics market and emerging applications (e.g., virtual and augmented reality). New pixel designs are being developed (e.g., "meta-OLED" pixels in Won-Jae *et al*., 2020) leading to novel displays with ultra-high pixel densities (>10,000 pixels per inch, or 1 pixel every ~2um). Their fabrication relies on nanolithographic techniques which provide very accurate control over the placement of pixels. Assuming the same level of relative accuracy as the phone screen, these devices may achieve 5 microarcsecond astrometric accuracy (e.g. 0.1 mas • 560 ppi / 10000 ppi). Such promising devices should be characterized to assess their potential as astrometric calibrators.

## 5. Acknowledgements

The authors gratefully acknowledge support of this research by the Mt. Cuba Astronomical Foundation, grant #5306. We also thank the anonymous referee for comments which improved this manuscript.